# WAVE$^{T*}$, a custom device able to measure viscoelastic properties of wood under water-saturated conditions

*(\* WAVE$^T$ : Environmental Vibration Analyser for Wood)*

Vincent Placet, Joëlle Passard, Patrick Perré

Laboratoire d'Etude et de Recherche sur le MAtériau Bois

LERMAB-UMR INRA/ENGREF/UHP 1093

ENGREF 14, rue Girardet 54042 Nancy cedex, France

Tel: 33 (0)3 83 39 68 92

Fax: 33 (0)3 83 39 68 47

E-mail: perre@nancy-engref.inra.fr

**Abstract**

This work presents an original experimental device conceived to characterise the viscoelastic properties of wood. Classically, the dynamic mechanical analysis of wood is performed using a commercial apparatus like a DMA (Dynamic Mechanical Analyser). However, when analysing wood with this type of apparatus, many problems related to the hygroscopic behaviour and the orthotropic structure of wood may be encountered. This is why an original apparatus perfectly adapted to the wood features has been developed. The WAVE$^T$ is able to measure the viscoelastic properties of wood samples under water-saturated conditions, in the temperature range of 5°C to 95°C at frequencies varying between 0.005 Hz and 10 Hz. Samples are tested in a cantilever configuration. The whole experiment has been designed to withstand the severe conditions of temperature and humidity.

At the same time, an analytical model based on Kelvin's elements has been developed. This model is able to correct experimental measurements performed close to the resonance frequency.

Results obtained for beech samples in radial and tangential directions using the WAVE$^T$ and a commercial apparatus (DMA 2980 TA Instruments) are compared and discussed. This comparison underlines the relevance of the WAVE$^T$ device.

<u>Keywords</u>: Experimental device / Viscoelasticity / WAVE$^T$ / Wood

# 1. Introduction

The viscoelastic behaviour of wood is essential in many fields of the wood industry. It is particularly important in wood processing operations such as drying, wood forming, thermal treatment, panel pressing, wood gluing and so on… This behaviour depends on the wood species (Hamdan *et al.* 2000, Olsson and Salmén 1992, 1997), the material direction and the type of wood: juvenile wood, sapwood and reaction wood (Lenth 1999, Lenth and Kamke 2001). Moreover, temperature and moisture content levels dramatically affect the viscoelastic behaviour of wood (Goring 1963, Ranta-Maunus 1975, Irvine 1984, Takahashi *et al.* 1998, Obataya *et al.* 1998, Olsson *&* Salmén 1997, Matsunaga *et al.* 2000).

During the second part of the last century, the rheological properties of wood have been widely studied thanks to various experimental methods. In the softening area, many properties such as mechanical, thermal, electrical, magnetic properties noticeably change.

For a long time, due to its simplicity, the dilatometric method was the only method used to determine the glass transition temperature Tg. Salmén (1990) accomplished many works in this domain. Some authors (Irvine 1984, Ostberg *et al.* 1990, Kelley *et al*, 1987) used DSC (Differential Scanning Calorimeter) or DTA (Differential Thermal Analysis). Indeed, specific heat capacity variation is actually an excellent indicator of transitions. Dielectric Thermal Analysis (DETA) is also quite a common method to explore molecular phenomena like relaxation (Lenth and Kamke 2001, Suzuki and Kuroda 2003, Maeda and Fukada 1987).

Several mechanical experiments have been used to characterise the viscoelastic properties. Among these, we often discern quasi-static methods: creep (Genevaux 1989, Guerrin 1990, Le Govic 1992, Perré and Aguiar 1999, Passard and Perré 2001, 2005b) and relaxation (Takahashi *et al.*, 1998) from harmonic and dynamic analysis. In comparison with creep tests, harmonic tests represent a more efficient method to characterise viscoelastic properties and glass transitions of materials. Their major advantage is to cover several decades of time independently of temperature control. In addition, this method allows the viscoelsatic behaviour to be distinguished from other effects such as ageing or growth stress recovery.

The typical frequency range of this forced non-resonance technique is 0.01 to 100 Hz. Because the experiments occur below the fundamental mode of vibration, the upper bound of the frequency range depends upon the sample stiffness and geometry. Nowadays, numerous commercial apparatus covering wide ranges of frequencies.

Other dynamic mechanical methods are available, including ultrasound and a variety of resonant techniques. The free-free flexural method exploits the frequency resonance of materials. This method is widely used, for example, to measure the wood characteristics for musical wind instruments (Matsunaga *et al.* 2000, Obataya *et al.* 2001, 2003, Lord 2003). At high frequencies, the methods are based on ultrasonic wave propagation. However this type of experiment is used in the wood field to measure stiffness rather than to determine internal friction (Bucur and Archer 1984).

The most common methods applied to wood at low and medium frequencies use commercial devices, like DMA (Dynamic Mechanical Analyser) or DMS (Dynamic Mechanical Spectroscope), (Kelley *et al.* 1987, Olsson and Salmén 1992, Björkmann and Salmén 2000, Obataya *et al.* 2003). However, large differences may be observed between data obtained by different commercial instruments (Hagen *et al.* 1994). A number of mechanical factors, like mechanical inertia (resonance frequency), specimen geometry and size, clamping effects may influence the results of dynamic mechanical analysis. Actually, all modes of geometry (flexion, traction, shear, compression…) proposed by commercial apparatus exhibit bias due to imperfect clamping and sample geometry. To make up for this lack of precision, the DMA technology applies correction factors to determine the viscoelastic properties of samples. These corrective factors are fitted for isotropic material, because this type of device is intended for classical polymers.

In comparison to these products, wood presents specific features, namely its anisotropy and its hygroscopicity. The former is problematic concerning the corrections factors and the latter is particularly important because water is recognised to act as a very efficient plasticisers in wood. Therefore, it absolutely requires to control the humidity during dynamic mechanical analysis.

Very few DMA apparatus do have humidity control, and even so, it is very difficult to keep the same moisture content of wood when the temperature changes. Notice that the moisture content variations, like drying, may cause collapse and mechano-sorptive effects.

For all these reasons, the wood rheology team at LERMAB developed an original experimental device, WAVE$^T$, able to perform dynamic mechanical tests of wood under water-saturated conditions and high temperature. WAVE$^T$ stands for Environmental Vibration Analyser for Wood. The working principle of the WAVE$^T$ consists in applying a harmonic force and to measuring displacement independently. Due to the time-dependent properties of polymers, the displacement response is out-of-phase with the applied stimulus. Thus, the storage modulus E' is defined as the in-phase or elastic response, proportional to the recoverable or stored energy; and the loss modulus E'' is the imaginary or viscous response, proportional to the irrecoverable or dissipated energy. In this forced non-resonance technique, samples are tested in a bending configuration, the single cantilever mode. In addition, great care was taken to grab, treat and analyse raw data.

## 2. Description of the experimental device

The WAVE$^T$ is dedicated to the measurement of stiffness and damping properties of water-saturated wood samples subjected to a deformation by periodic stress. More specifically, this device is able to determine the evolution of the viscoelastic properties of green wood samples with regard to temperature: 5 to 95 °C, frequency: $5.10^{-3}$ to 10 Hz and stress level: 0.01 to 4 MPa. To maintain sample water saturation during experimentation and to avoid any mechanosorptive effects due to temperature gradients, samples are held in a temperature-controlled water bath.

The WAVE$^T$ is composed of a complete chain of vibration analysis (1) and of a test bench (2) (Figure 1). It includes a conditioning chamber (3) which allows the immersion and the temperature control of the wood sample during experimentation.

### 2.1 Principle

A harmonic force is applied to the sample using a mini-shaker. The mini-shaker is of the electro-dynamic type with a permanent field magnet (Brüel & Kjaer type 4810). An alternating voltage signal provided by a function generator is converted into current signal by an operational amplifier and sent to the magnetic coil of the mini-shaker. The latter produces a sinusoidal movement transmitted to the sample via a light-weight stainless-steel tube.

In most commercial instruments, the force applied to the sample is deduced from the input signal to the electro-magnetic coils in the driver. In our device, a miniature load sensor cell is used to measure the applied load. This sensor is integrated into the excitation bar, between the sample and the mini-shaker.

Typically, this type of mini-shaker is conceived to work in a frequency range of 1 Hz to 18 kHz. Below 1 Hz, the mechanical excitation applied to the sample loses its sinusoidal shape, due to the suspension and the rubber protection film of the driver. To address this problem, a feedback loop was built with suitable PID parameters, so the force applied to the sample perfectly follows the generator signal. In this way, it is possible to apply the desired sinusoidal force at very low frequency levels: tests have been performed successfully at 1 mHz, but the only limitation is indeed the time required to get experimental data.

In order to avoid any disturbance at the clamps area, and contrary to classical DMA, the deflection measurement is dissociated from the stress system (Fig. 1).

Force and displacement signals are treated by electronic conditioners (filters and amplifiers) and captured by a multichannel digital oscilloscope (Agilent 54624A). The megazoom technology of this apparatus permits us to follow the evolution of the signals and particularly the phase difference over large frequency and amplitude ranges.

The temperature is measured by a J-type thermocouple placed close to the sample. To keep the sample saturated, tests are conducted in water. Consequently, the raw experimental data integrate the phase difference due to the wood viscosity but also the phase difference due to water viscosity. Measurements carried out on a pure elastic sample (metal) in different environmental conditions (air, water and water in circulation) revealed the influence of water, especially for frequency levels higher than 1 Hz, but only as a random noise. No systematic damping effect was observed over our range of frequency. This noise is probably due to the water circulation. It is significantly reduced by using several periods of the sine curves to determine the viscoelastic parameters (for example, 20 periods are used at 10 Hz).

### 2.2 Deformation mode and mechanical configuration

The single cantilever configuration is depicted in figure 2. One end of the sample is tightly clamped (in our tests, the sample thickness is always along the very rigid longitudinal direction) and the harmonic force is applied on the other end. The beam theory, which assumes some hypotheses (small perturbation , Navier-Bernouilli, Saint-Venant…), allows the deflection to be obtain as a function of the deformation mode, the force level, the sample length and the location where the displacement is measured :

$$H_{bend} = \frac{Fl_0^2(6L - 2l_0)}{bh^3 E} \qquad (1)$$

where $F$ is the sample load (N), $L$ the sample length (m), $l_0$ the distance to the deflection measuring point (m), $b$ the beam width (m), $h$ the beam thickness (m) and $E$ the modulus of elasticity (Pa).

The previous equation accounts only for the pure bending mode of the beam. The influence of shear stress on the deflection can be evaluated by the following formula, proposed by Timoshenko (1968):

$$H_{tot} = H_{bend}\left(1 + \alpha \frac{h^2}{l_0(6L - 2l_0)}\frac{E}{G}\right) \qquad (2)$$

G is the shear modulus (Pa) and $\alpha$ is a coefficient which depends on the specimen shape. In the case of single cantilever bending and for a parallelepiped beam, $\alpha$ is equal to 3/2.

Equation (2) shows that the deflection due to shear stress depends directly on the sample geometry. The thinner the sample and the further the deflection measurement point is from the fixed fitting, the less the influence of shear stress is perceptible. In addition, in a case of an anisotropic material such as wood, the ratio E/G depends on the material direction. This ratio is weaker in transverse directions that in longitudinal direction. Then, considering material directions (radial and tangential directions) and the shape of the specimen ($L > 10h$), the effect of shear stress remains negligible (Passard and Perré 2005a).

### 2.3 Test bench

The device is put on an anti-vibratory table, to limit the disturbances due to the external vibrations. The bottom part of the vibratory system holds the sample holder. The top part supports the force and displacement sensors and the shaker. The two parts are separated by an isolating plate which protects the electronic apparatus from temperature and moisture. To avoid any heat bridges, Nylon screws are used to connect both parts. The loading tube crosses the plate through a small hole (Figure 3.a-b).The system is hung on a crossbar through an electric jack, which permits the system to be vertically moved : immersion of the sample holder in the measuring position, accessibility of the sample holder in the handling position.

<u>Sample holder</u>

A specific articulated clamp was designed to apply a pure vertical force without any momentum to the sample and without any vertical clearance (Figure 3.c).

The system is made up of two parallel clamps (2) fixed to a support (1) through a shaft (6) supported by four ball bearings (7-8). The assembly of the ball bearings is relatively subtle. At each extremity of the shaft, the ball bearings are doubled. One is fitted (8) and the other is assembled with a preload (7). In this way, the clamp presents no radial clearance but allows thermal extension without locking of the balls. Both conditions were required to perform tests at zero mean load over a wide range of temperature. To avoid any induced momentum to the beam, the shaft is situated on the neutral line of the sample. The whole clamping system is made of stainless steel.

On the other side, the sample is fitted between two plates, compressed with four screws and springs. Each screw is tightened with a constant torque. The springs (4) allow the potential clearance due to thermal dilatation and relaxation of wood to be balanced.

In equation (1) the clamp is considered as perfect. In our configuration, this assumption is reasonable because the sample is compressed in the longitudinal direction. Therefore, the modulus of elasticity is much lower in the measuring direction than in the clamping direction.

Force measurement

The force sensor used is a miniature load sensor (FGP Instrumentation type XFTC 300). This is appropriated to measure tension and compression in static and dynamic tests. The sensing element is fitted with a fully temperature compensated Wheatstone bridge equipped with high stability strain gauges micro-machined on a monocrystal of silicon. Such a feature limits considerably the sensor shift over long periods of time The full-scale range is 0-20 N. Although this material is miniature, it is robust and presents a high stiffness.

Deflection measurement

Considering the difficulties presented in carrying out dynamic deflection measurements in water saturated and aggressive environments at high temperature, the displacement measurement system is transferred out of the conditioning room. The system is composed of a double-groove pulley, a spring, a cable and a minute stirrup tied to the sample (Figure 3).

The pulley axis is articulated on ball bearings. A cable tied to a spring is wound around the small-diameter groove of the pulley. Another stainless steel cable is wound around the large-diameter groove of the pulley in the opposite direction, and is tied to the sample using the stirrup. So, the vertical motion of the sample is converted into a rotation of the pulley. A laser micrometer (Bullier type M5L/4) read the displacement at the diameter of the pulley. Its full-scale range is +/- 2 mm with a resolution of 1 $\mu$m, using an integration time consistent with the maximum frequency used during our tests. The signal distortion due to the angle effect of the rotation can be easily corrected using trigonometric functions and is taken into account in our calculations.

The stirrup is equipped with two metal spikes, which prevent it from sliding on the sample during the experiment. These spikes also permit the stirrup to keep its vertical orientation in spite of the rotation of the sample surface due to bending.

# 3. Analysis of raw data

## 3.1 Determining the phase difference

The main characteristic of harmonic tests is the sinusoidal form of force and displacement functions. Both signals have the same frequency but are out of phase. This phase difference δ is precisely the key parameter that allows the viscoelastic behaviour of the sample to be calculated. Therefore, the relevance of the measurements depends on how accurately the phase difference can be obtained. In particular, the method should be able to deal with the experimental noise level. This is why an inverse method was chosen to extract this phase difference over several periods of the sinusoidal signals.

The raw signals of force and displacement, grabbed by the oscilloscope, are downloaded on a computer by RS232 connectors. The theoretical equations of force and displacement curves are:

$$F = Y_1 + F_0 \times \sin(2.\pi.f.t + \alpha_1) \tag{3}$$

$$H = Y_2 + H_0 \times \sin(2.\pi.f.t + \alpha_2) \tag{4}$$

$$\delta = \alpha_1 - \alpha_2 \tag{5}$$

where $F$ is the force (N), $H$ the displacement (mm), $F_0$ the force amplitude (N), $H_0$ the displacement amplitude (mm), $Y_1$ and $Y_2$ the average force and displacement, $f$ the frequency (Hz), $t$ the time (s) and $\alpha_1$ and $\alpha_2$ the signal phases.

The inverse method needs an objective function φ to evaluate the discrepancy between experimental and theoretical curves. We simply use the Euclidian norm :

$$\varphi = \frac{\sum_n (F_{calculated} - F_{measured})^2 + \sum_n (H_{calculated} - H_{measured})^2}{n} \tag{6}$$

with n, the number of experimental points.

This function depends on 7 parameters ($Y_1$, $Y_2$, $F_0$, $H_0$, $\alpha_1$, $\alpha_2$, $f$) which are identified by minimising the function φ using the simplex algorithm (Press *et al.* 1992).

This procedure assumes that the solution is the global minimum. Nevertheless, the risk of falling into local minima is important. To avoid this, the initial estimated values must to be close enough of the solution. To insure this, initial parameters are calculated on each individual curve for 3 parameters (average, amplitude and phase), before the identification is performed on both curves using 7 parameters. Identifying the frequency in this final step proved to be required when several periods are used to reduce the noise level: in this case, a very small mistake in the frequency value would spoil all the identification procedure. The stability of the solution is also tested afterwards by introduction of the optimised values as initial values, with a large initial search simplex. Finally, the values are accepted or rejected according to the final value of φ.

### 3.2 Model

Throughout the development of the device, several precautions were taken to ensure the relevance of the experimental data. The different apparatus functions (precision, calibration, resonance frequency…), the linearity of the viscoelastic behaviour, the stability and the reproducibility of measurements for reference materials (metal)… were carefully checked.

At the same time, to correct or reject measurements done close to the resonance frequency, an analytical model was developed. It allows the phase change due to the inertia of all moving parts, the damping effects (water and rubber membrane of the shaker) and the frame stiffness to be dissociated from the sample properties.

The frame stiffness has been determined using a very stiff sample (rather thick beam of stainless steel). The experimental deflection measured in this test allowed the frame compliance to be determined ($5.9.10^{-3}$ mm.N$^{-1}$).

Concerning the inertia effect, it is classically considered that measurements are valid up to one tenth of the frequency resonance value, which depends on the stiffness and the geometry of the sample. For example, for a typical wood sample, if the resonance appears near 30 Hz, the measurements are only valid below 3 Hz (one tenth). However, the model enables data to be corrected up to 10 Hz (one third of the resonance frequency).

In figure 4-a, the loss factor of a metal sample is measured at different frequencies. Consistently, the raw tanδ value drops near the resonance (50 Hz). The ability of the model to correct this effect is obvious on this graph.

The model can also be used to predict the behaviour of a sample submitted to harmonic tests. In particular, it allows the resonance frequency to be rather well predicted (Figure 4-b). This graph depicts the measured and the simulated deflection amplitude around the resonance. In the model, the sample properties measured before the resonance are used as input parameters. The slight discrepancy observed between model and experiment is due to:
- the resonance itself, which perturbs the experimental signals especially near the peak of the curve,
- the assumption of constant viscoelastic properties, which should rigorously depend on the frequency (right hand side of the curve).

### 3.3 Device control

The functioning principle of the vibratory analysis chain is summarized in figure 5. All these operations are controlled by a software developed in Visual Basic. The program communicates with the different apparatus and uses a DLL module (Dynamic Link Library) written in Fortran to identify the parameters. This application also includes the analytical model, so that only corrected values are stored on the disk. This ensures also that any measurement done too close from the resonance frequency will be automatically rejected. The operator has just to key in the desired experimental protocol (frequency, temperature stages, stress level…) and to retrieve the file containing the treated data at the end of the experiments. The pilot program sets the thermocryostat at the required temperature levels, produces the force and displacement measurements once the sample temperature is stabilised, identifies the phase difference value using the simplex algorithm extract the corrected data as given by the device model and repeats this process for all the requested frequencies and temperature stages. The values are accepted or rejected according to the criterion. The treated data are synthesized in a file, with graphs describing the evolution of the storage modulus, the loss modulus and the loss factor as a function of frequency and/or temperature.

The first results obtained for wood using the WAVE$^T$ depicts the evolution of the storage modulus and the loss factor as a function of temperature and frequency, for an oak sample in the radial direction (Fig. 6). For a constant frequency, the storage modulus decreases with increasing temperature. Consistently, the curves of the storage modulus are perfectly set out in tiered rows, with a storage modulus at a given temperature that increases with frequency. This property is shared by all viscoelastic solid materials, and finds its theoretical explanation in the Kramers-Krönig relations.

In figure 6 the tanδ peaks, expressing a relaxation of wood, are attributed to the transition of lignin (Goring 1963, Olsson and Salmén 1992, Hamdan *et al.* 2000). The softening temperature (Tg) shifts to higher values as the frequency increases and the maximum value of the loss factor increases with frequency.

These trends are in a good agreement with literature data (Salmén 1984). Therfore, these first results are consistent and encouraging. For further positive feedback, it should be rewarding to compare the measurements gathered with the WAVE$^T$ disposal to those of commercial apparatus.

## 4. Comparison with commercial apparatus

A comparative study was preformed to compare results obtained with the WAVE$^T$ and a commercial DMA (2980 TA Instruments). The DMA does not have a conditioning chamber with a water bath or a humidity controller. To keep the samples saturated during the tests, they are wrapped up with cellophane. This method is relatively effective, as we observe that the water loss remains negligible during the tests (less than 3 or 4 % of loss in mass, which only affect the liquid water).

### 4.1 Wood sampling

Variability is an extremely important characteristic of wood. Because the samples tested with the two apparatus need to be as similar as possible, the sampling was crucial in this study. Beech wood (*Fagus sylvatica*) has been chosen for this comparison because this is a pore diffuse species: its anatomical structure is almost the same all over the entire annual growth ring.

Two series of samples were cut from the heartwood of the green log of a 60-year old beech tree, the first in the radial direction, the second in the tangential direction. Each series was prepared from the same annual growth rings, next to the other, along the longitudinal direction. Table 1 shows that the properties, and particularly the storage modulus, is almost constant within a series of samples. The maximum deviation is 1% for the density and 5% for the modulus, with is excellent for a material of biological origin.

Once cut, the samples are immediately soaked in water in an airtight box and put in a refrigerator to limit fungal attacks. Then, samples are tested within a few days. The sample section is 5×10 mm², its length is 30 mm for the DMA and 100 mm for the WAVE$^T$.

### 4.2 Experimental procedure

For each material direction, five samples are tested with each apparatus. There are submitted to multifrequency temperature scans, for a maximum stress levels of about 0.6 MPa.

With the WAVE$^T$, measurements are performed every 5°C between 5°C to 95°C. The heating rate is 0,35°C.min$^{-1}$. At each step, the temperature of the water bath is maintained during 2 minutes to stabilise the sample temperature before measurements.

The accuracy of the WAVE$^T$ and the reproducibility has been tested using standard samples of polymer (Placet 2006)

The same procedure is used for the DMA. The temperature range is 35°C to 95°C (no cooling option available on this DMA), the heating rate 0,25°C. min$^{-1}$, and the isothermal plateau duration of about 1 minute. Measurements are carried out every 2°C.

At each temperature step, a frequency scanning is done (0,1 – 1 – 10 Hz) and for each frequency, the viscoelastic properties (E', E'' and tan$\delta$) are determined. The samples are tested in single and dual cantilever bending mode respectively for the WAVE$^T$ and the DMA.

### 4.3 Results and discussion

Table 2 and figure 7 show that the storage modulus is always smaller, by 25% to 30%, with the WAVE$^T$ than with the DMA, regardless of the material direction and the frequency. The anisotropy ratio ($E'_R/E'_T$), however, is similar with the two apparatus: 1.43 for the WAVE$^T$ and 1.56 for the DMA.

The trend of the DMA to get stiffer samples may be explained by the geometrical factor. The dual cantilever bending mode proposed by the DMA leads to a very short sample length: its actually simulates two 8-mm long single cantilever samples in series, against one 60-mm long sample for the WAVE$^T$. Such short samples do not respect the hypothesis of the beam theory. Consequently, a correction factor was implemented in the device, but the latter was calculated by finite element methods for isotropic materials, which is not suitable for wood.

Hagen *et al.* (1994) already underlined this problem with this type of apparatus. They showed, on natural rubber, that the incorrect sample geometry required by the apparatus leads to results indicating a stiffer material than in reality. They also emphasised, that it is possible to record differences as high as 14°C in the determination of Tg with two different apparatus, certainly due to temperature calibration or sensor location.

The evolution of the tanδ (figure 7) exhibits a softening temperature of beech between 80 and 90°C at a frequency of 1Hz, corresponding to the softening of saturated lignin. The shape of the curves near 5°C obtained with the WAVE$^T$ suggests a possible existence of a second tanδ peak below 0°C. According to Olsson and Salmén (1997) and Lenth (1999) this may be due to the relaxation of hemicelluloses.

The two apparatus detect a higher value of the loss factor in the tangential direction than in the radial direction (at least at the peak level). Contrary to the DMA, which finds a softening temperature almost identical in the radial direction and the tangential direction, the WAVE$^T$ shows a noticeable difference. The softening temperature in the radial direction (82°C) is lower than in the tangential direction (near 90°C). The only paper we found which differentiates the two transverse directions (Backman and Lindberg 2001) confirms these trends and seems consistent with our custom device.

## 5. Conclusion

In this work, an original experimental device devoted to the dynamic mechanical analysis of wood (hygroscopic and orthotropic heteropolymer) is presented. The WAVE$^T$ is capable of automatically performing multifrequency tests (0.005 to 10 Hz) on green wood samples up to 100°C. Several precautions have been taken to ensure the relevance of the device. An analytical model was developed to dissociate the phase change due to the inertia of all moving parts, the damping effects (water and rubber membrane of the shaker) and the frame stiffness from the sample properties.

Our first results are consistent with literature data. By comparison with a standard DMA, our custom device brings several improvements: actual measurement of the force and deflection signals, load-free deflection measurement, rigorous sample geometry, perfect saturation conditions, rigorous single cantilever mode,

Henceforth, we have at our disposal an excellent tool to study wood according to the material direction, the species, the type of wood (clones, reaction wood), the chemical or thermal treatment...

We have also to mention that this custom device is a flexible tool that can be adapted to specific configurations. In the future, two main innovations will be implemented : i) a pressure chamber to perform tests with soaked samples over the boiling point of water ii) a controlled gas supply, based on a mixture of dry gas and saturated gas (Perré *et al.* 2007), to accurately control the relative humidity and their variations during the tests.

## 6. References

Backman, A.C.; Lindberg, K.A.H., 2001. Differences in wood material responses for radial and tangential direction as measured by dynamic mechanical thermal analysis. Journal of Materials Science 36, 3777-3783.

Björkmann, A. and Salmén, L., 2000. Studies on solid wood. II. The influence of chemical modifications on viscoelastic properties. Cellulose chemistry and technology 34, 7-20.

Bucur, V.; Archer R.R., 1984. Elastic constants for wood by an ultrasonic method. Wood Science and Technology 18, 255-265.

Genevaux, J-M., 1989. Le fluage à température linéairement croissante : caractérisation des sources de viscoélasticité anisotrope du bois. PhD thesis, INPL, Nancy, France.

Goring, D.A.I., 1963. Thermal Softening of Lignin, Hemicellulose and Cellulose. Pulp and Paper Magazine of Canada, 517 -527.

Guerrin, G.M., 1990. Caractérisation en flexion quasi-statique et dynamique d'un matériau thermo-hygro-visco-élastique : le bois. PhD thesis, INPL, Nancy, France.

Hagen, R. ; Salmén, L. ; Lavebratt, H. and Stenberg, B., 1994. Comparison of dynamic mechanical measurements and Tg determinations with two different instruments. Polymer Testing 13(2), 113-128.

Hamdan, S. ; Dwianto, W. ; Morooka, T. and Norimoto, M., 2000. Softening Characteristics of Wet Wood under Quasi Static Loading. Holzforschung 54(5), 557-560.

Irvine, G.M., 1984. The glass transitions of lignin and hemicellulose and their measurement by differential thermal analysis. Tappi Journal 67(5), 118-121.

Kelley, S.S. ; Rials, TG. and Glasser, WG., 1987. Relaxation behaviour of the amorphous components of wood. Journal of Materials Science 22, 617-624.

Le Govic, C., 1992. Panorama des travaux scientifiques mondiaux sur le fluage du bois sans défaut (effet de la température et de l'humidité). CTBA, Paris, 75-82.

Lenth, C.A., 1999. Wood material behavior in severe environments. PhD thesis Wood Science and Forest Products, Virginia Polytechnic Institute Blacksburg.

Lenth, C.A.; Kamke, F.A., 2001. Moisture dependent softening behavior of wood. Wood and Fiber Science 33(3), 492-507.

Lord, A. E., 2003. Viscoelasticity of the giant reed material Arundo donax. Wood Science and Technology 37, 177-188.

Maeda, H. and Fukada, E., 1987. Effect on Bound Water on Piezoelectric, Dielectric, and Elastic Properties of Wood. Journal of Applied Polymer Science 33, 1187-1198.

Matsunaga, M. ; Obataya, E. ; Minato K. and Nakatsubo F., 2000. Working mechanism of adsorbed water on the vibrational properties of wood impregnated with extractives of pernambuco. Journal of Wood Science 46, 122-129.

Obataya, E.; Norimoto, M. and Gril, J., 1998. The effects of adsorbed water on dynamic mechanical properties of wood. Polymer 39(14), 3059-3064.

Obataya, E.; Minato, K. and Tomita, B., 2001. Influence of moisture content on the vibrational properties of hematoxylin-impregnated wood. Journal of Wood Science 47, 317-321.

Obataya, E.; Furuta, Y. and Gril, J., 2003. Dynamic viscoelastic properties of wood acetylated with acetic anhydride solution of glucose pentaacetate. Journal of Wood Science 49(2), 152-157.

Olsson, A.-M. and Salmén, L., 1992. Chapter 9 : Viscoelasticity of In Situ Lignin as Affected by Structure. Softwwood vs. Hardwood. Viscoelasticity of Biomaterials. American Chemical Society, 133-143.

Olsson, A.-M. and Salmén, L., 1997. Humidity and temperature Affecting Hemicellulose Softening in Wood. International Conference of COST Action E8. Mechanical Performance of Wood and Wood Products. Wood-water relation. Copenhagen, Denmark.

Ostberg, G. ; Salmén, L. and Terlecki J., 1990. Softening temperature of Moist Wood Measured by Differential Scanning Calorimetry. Holzforschung 44(3), 223-225.

Passard, J. and Perré, P., 2001. Creep tests under water-saturated conditions: do the anisotropy ratios of wood change with the temperature and time dependency ? 7th International IUFRO Wood Drying Conference, Tsukuba, Japan, 230-237.

Passard, J. and Perré, P., 2005a. Viscoelastic behaviour of green wood across the grain. Part I Thermally activated creep tests up to 120°C. Ann. For. Sci., 62, 707-716.

Passard, J. and Perré, P., 2005b. Viscoelastic behaviour of green wood across the grain. Part II A temperature dependent constitutive model defined by inverse method. Ann. For. Sci., 62, 823-830.

Perré P. ; Houngan A.C. ; Jacquin Ph., 2007. Mass diffusivity of beech determined in unsteady-state using a magnetic suspension balance, Drying technology, 25, 1341-1347.

Perré, P. and Aguiar. O., 1999. Creep at high temperature (120°C) of "green" wood and modelling by thermo-activated Kelvin's elements. Ann. For. Sci., 56, 403-416.

Placet, V. 2006. Design and development of an innovative experimental device for the characterisation of the viscoelastic behaviour and the thermal degradation of wood in severe conditions. PhD Thesis, UHP, Nancy, France.

Press W.H.; Teukolsky S.A. ; Vetterling W.T. and Flannery B.P., 1992. Numerical Recipes in Fortran. The art of scientific Computing, 2$^{nd}$ ed. Cambridge University Press, Cambridge, 402-406.

Press W.H. ; Teukolsky S.A. ; Vetterling W.T. and Flannery B.P., 1992. Numerical Recipes in Fortran. The art of scientific Computing, 2$^{nd}$ ed. Cambridge University Press, Cambridge, 36-42.

Ranta-Maunus, A., 1975. The viscoelasticity of wood at varying Moisture Content. Wood Science and Technology 9, 189-205.

Salmén, L., 1984. Viscoelastic properties of In situ lignin under water-saturated conditions. Journal of Materials science 19, 3090-3096.

Salmén, L., 1990. Thermal Expansion of Water-saturated Wood. Holzforschung 44, 17-19.

Suzuki, Y. and Kuroda N., 2003. Dielectric Properties of Wood with High Moisture Content. Mokuzai Gakkaishi 49(3), 161-170.

Takahashi, K. ; Morooka, T. and Norimoto, M., 1998. Thermal softening of wet wood in the temperature range of 0 to 200°C. Wood research 85, 79-80.

Timoshenko S.P., 1968. Résistance des matériaux, Bordas, Paris.

# TABLES

# FIGURES

| Sample | Basic density (kg.m$^{-3}$) | E' (MPa) |
|---|---|---|
| 1 | 591 | 420 |
| 2 | 596 | 440 |
| 3 | 592 | 450 |
| 4 | 591 | 450 |
| average | **592** | **440** |

**Table 1 -** Comparison of the storage modulus of 4 samples taken from a beech tree in the tangential direction. Storage modulus measurements are performed with the WAVE$^T$, at a frequency of 1 Hz and a temperature of 35°C.

| | **Storage modulus MPa** (average value) | |
|---|---|---|
| | **Radial** | **Tangential** |
| DMA | 812 | 519 |
| **WAVE$^T$** | 568 | 396 |

**Table 2 -** Average values of the storage modulus of beech samples according to material direction and the apparatus used (frequency: 0.1 Hz, temperature: 40°C).

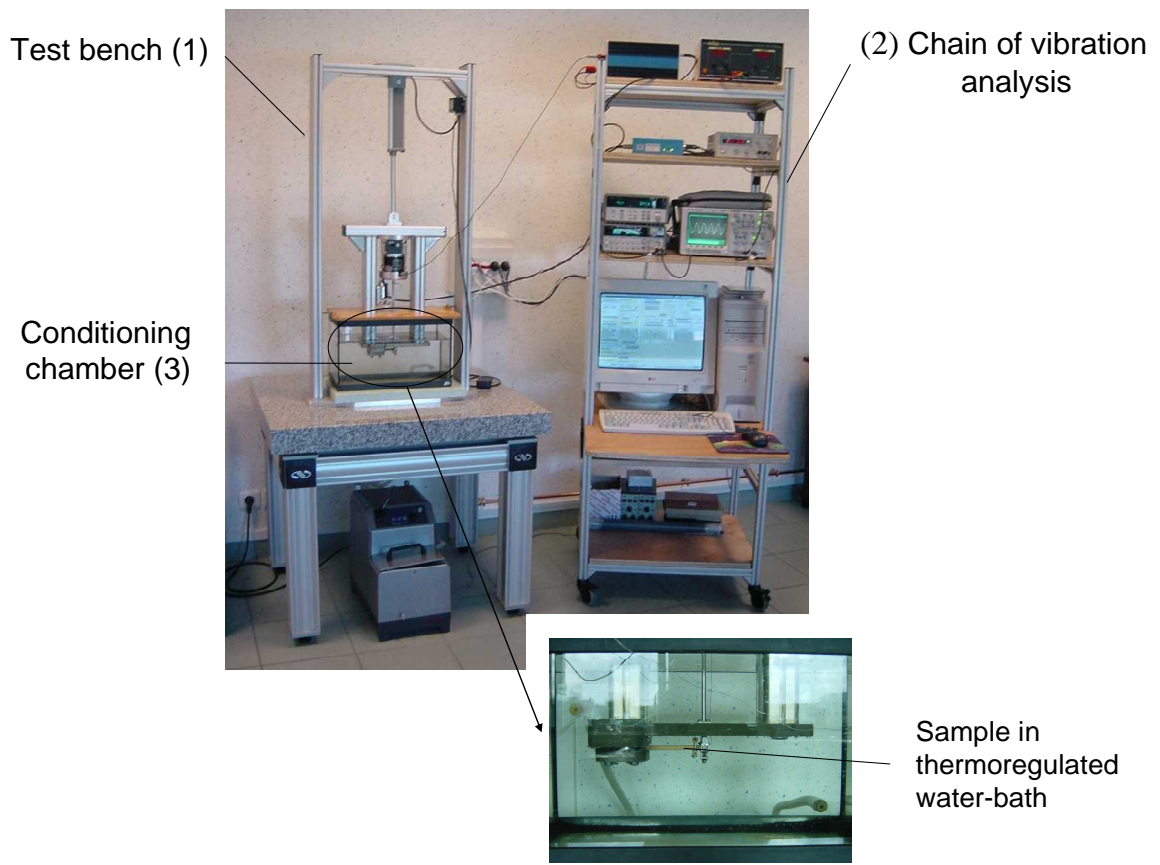

**Figure 1** – A overall view of the experimental device (the WAVE$^T$)..

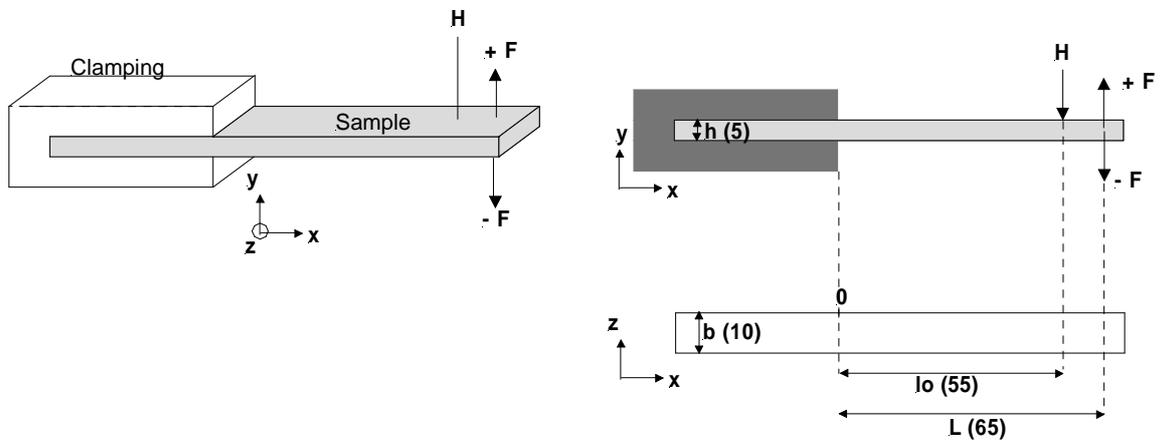

**Figure 2 -** Mechanical configuration (single cantilever bending) and sample geometry.

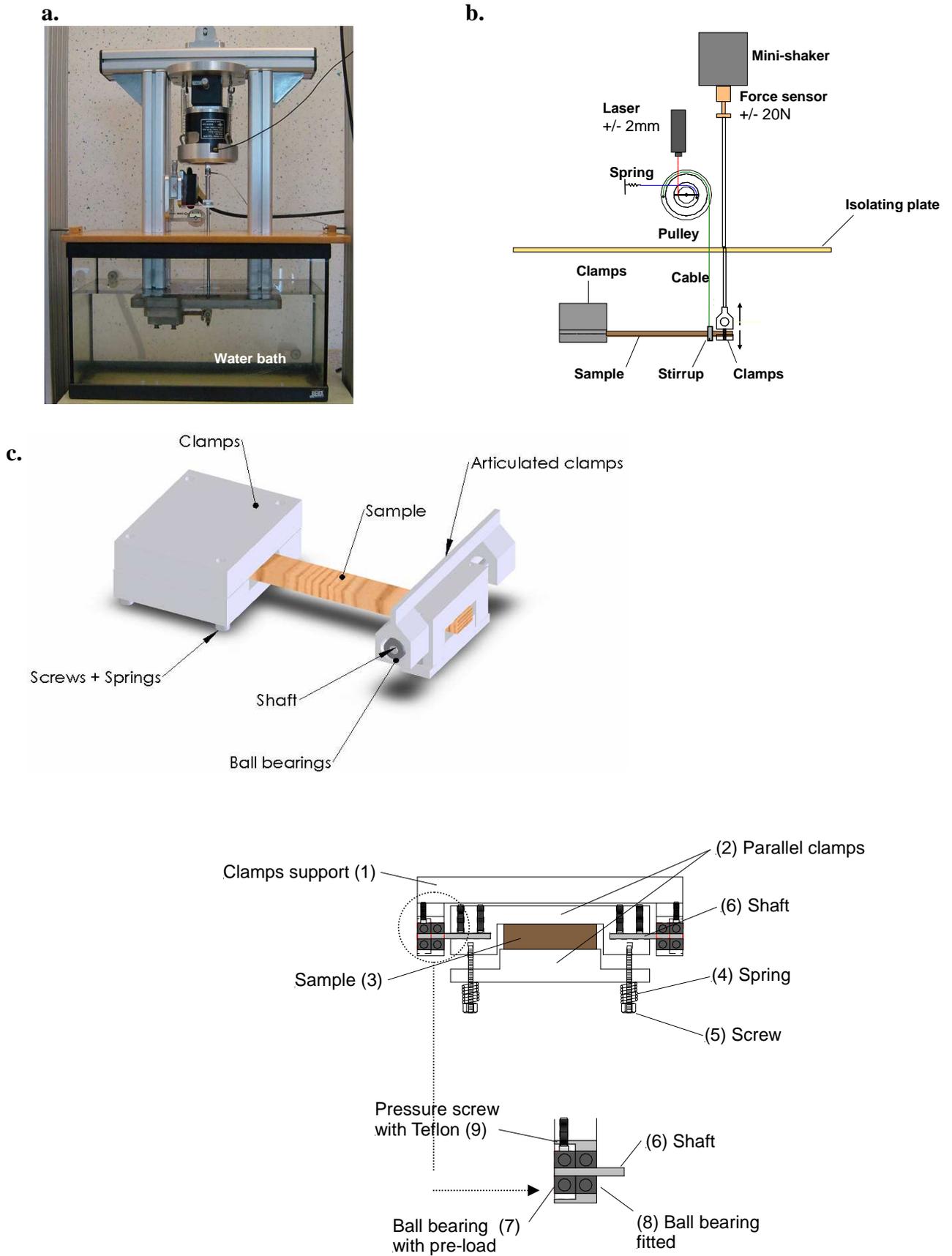

**Figure 3** – Details of the test bench. a. Photo of the test bench – b. Sketch of the deflection measurement and stress systems – c. Description of the clamps.

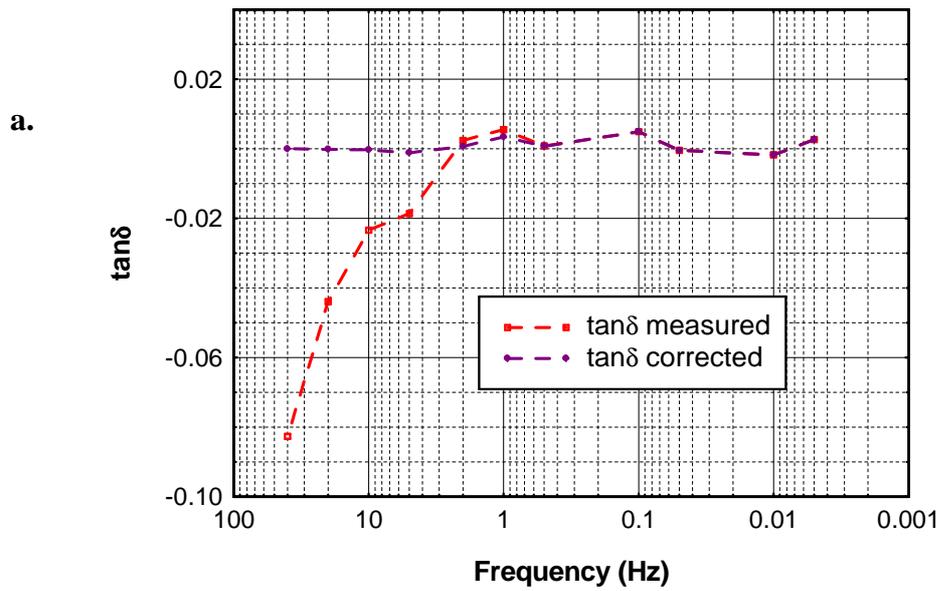

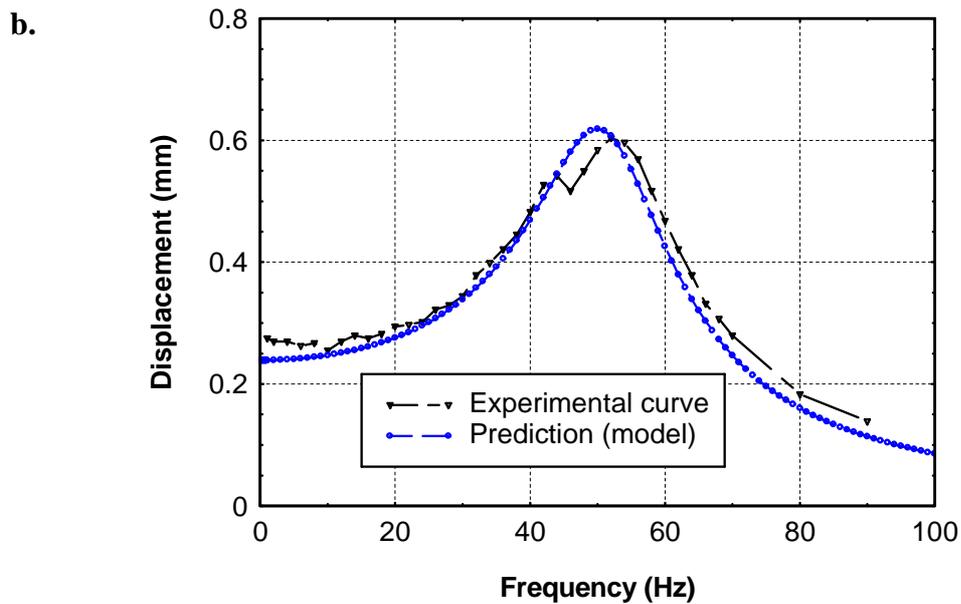

**Figure 4** - **a.** Using the model to correct the tanδ values at different frequencies for a pure elastic sample: brass (in this example, the resonance frequency was around 50 Hz). **b.** Comparison of deflection magnitude gathered by the WAVE$^T$ and value predicted with the model.

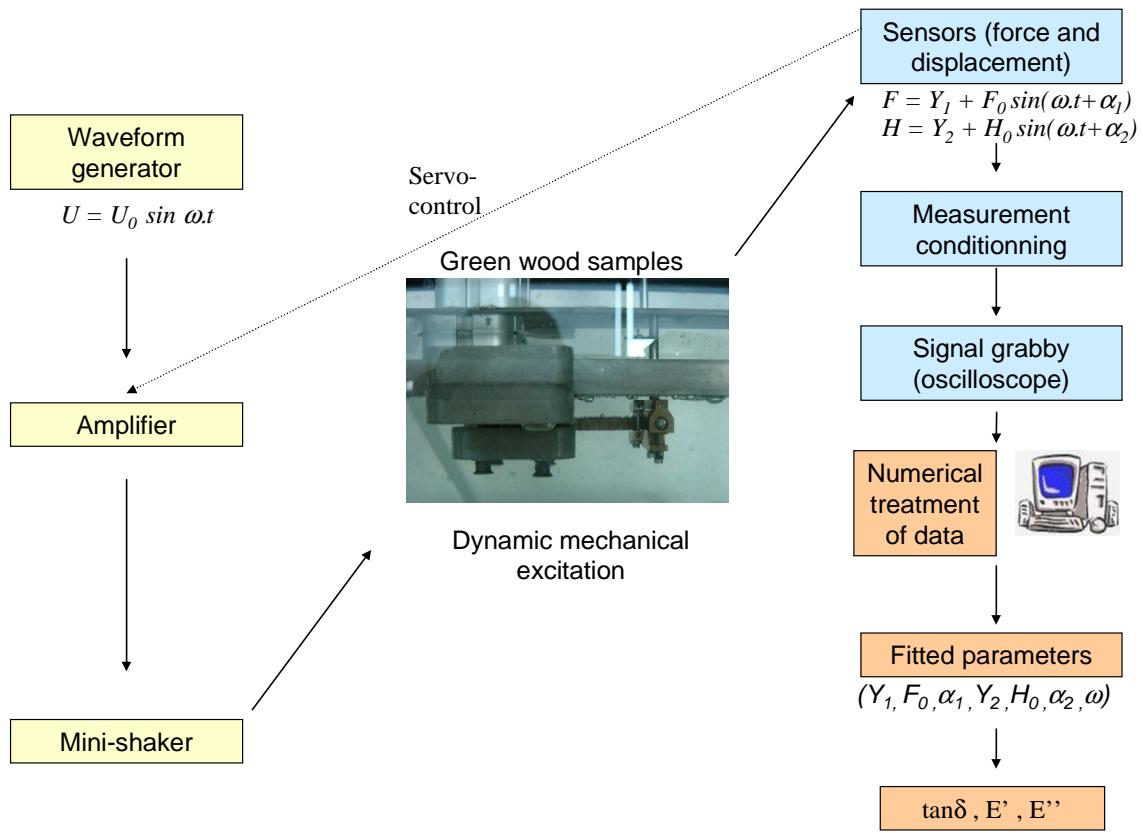

**Figure 5** – A schematic diagram of the vibratory analysis chain and signal treatment.

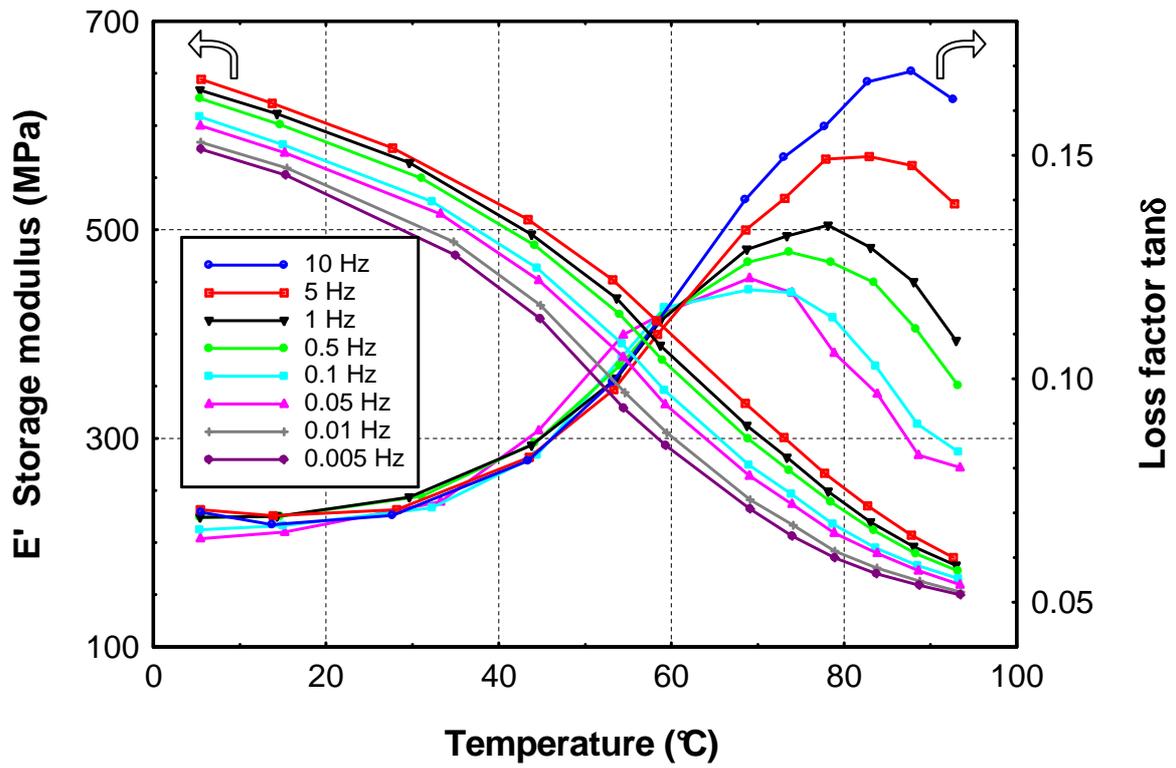

**Figure 6 -** Evolution of the storage modulus and the loss factor at different temperatures for several frequencies (Oak, radial direction).

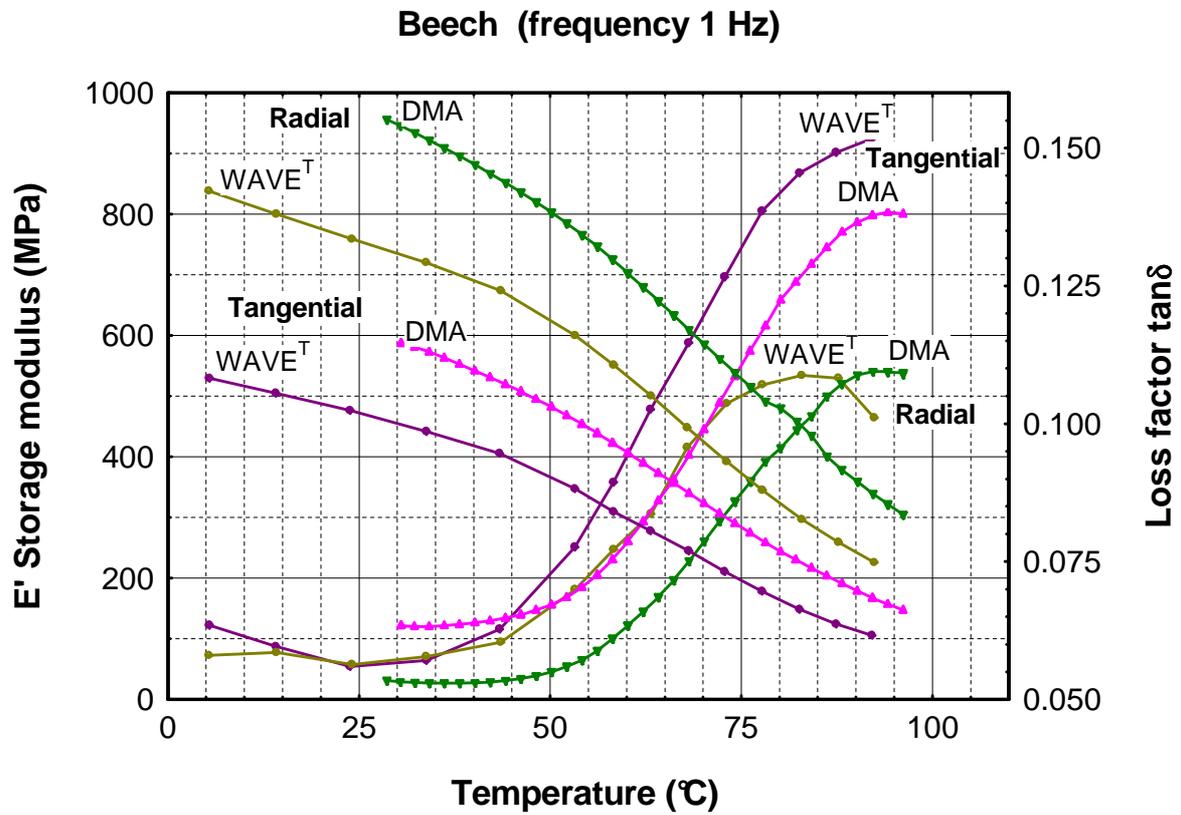

**Figure 7** - Evolution of the storage modulus and the loss factor with temperature: comparison of two apparatus (beech, tangential and radial direction, frequency = 1Hz)..